\documentclass[a4paper,10pt]{article}

\title{Constituent Quarks and the Spin of the Proton}
\author{H. Fritzsch and G. Eldahoumi  \smallskip  \\ \textit{Physics 
Department}\\ \textit{University of Munich, Germany}}

\usepackage{graphicx}

\begin{document}

\maketitle

\begin{abstract}
The constituent quarks are interpreted as bound states, which have an internal 
structure. The quark distributions of the proton are related to those of the 
constituent quarks. The experiments support this hypothesis. Likewise the spin 
structure of the proton is related to the spin structure of the constituent 
quarks. We find that about 30\% of the spin of a constituent quark is given 
by the valence quark, and 70\% are provided by the gluons. 
\end{abstract}
\bigskip
\bigskip 
In the Standard Model the strong interactions are described by the gauge theory 
of quantum chromodynamics, introduced in 1972 (ref.1,2). In this theory quarks 
and gluons interact and form the baryons and mesons as bound states. Both the 
massive quarks and the massless gluons are permanently confined and do not 
appear as free particles. 

The quarks were introduced in 1964 by M. Gell-Mann (ref.3) and G. Zweig (ref.4) 
as the constituents of the hadrons. In the deep inelastic scattering experiments 
of electrons and nucleons the quarks were observed indirectly at SLAC in 1968. 
But it was found that the nucleons consist not only of the three valence quarks, 
but also of many quark-antiquark pairs. 

Today we distinguish between constituent quarks and current quarks, but it 
remained obscure what role the constituent quarks play in the theory of QCD. 
In particular the magnetic moments of the baryons in the baryon octet are 
described very well  by the three constituent quarks, if one uses the SU(6)-wave 
functions. For example, the ratio of the magnetic moments of the proton and 
neutron is predicted to be -3/2 (ref.5). The experiments give for this 
ratio -1.46, i.e. experiment and theory agree within 3\%. The magnetic moments 
of the other baryons in the SU(3)-octet agree also very well with the model, 
if one adjusts the ratio of the quark masses $m_s/m_u$. Thus the constituent 
quarks seem to exist inside the baryons and mesons. Various long range 
properties of the hadrons, not only the magnetic moments, can be described 
rather well by the model of the constituent quarks. 

Using the SU(6) wave functions for the nucleons, one can also calculate the 
ratio of the axial vector coupling constant , describing the beta decay, and 
the vector coupling constant :
\begin{eqnarray}
\frac{g_{A}}{g_{V}}=\frac{5}{3}
\end{eqnarray}
However the experiments give for his ratio:
\begin{eqnarray}
\frac{g_{A}}{g_{V}}=1.2695\pm 0.0029
\end{eqnarray}
The experimental value is about 24\% less than expected. The theoretical 
prediction follows, if the three quarks are in an s wave. If the space wave 
function has contributions from non-zero angular momenta (p-wave etc.), the 
ratio would be reduced. We shall assume that non-zero angular momenta are 
present and lead to the reduced value of the axial vector coupling. In this 
case the predictions for the ratios of the magnetic moments would still be valid. 

We suggest that a hadron can be viewed as a loosely bound system of constituent 
quarks, which have an effective mass of about 330 MeV. However the constituent 
quarks are not pointlike objects, but composite systems, consisting of a valence 
quark, surrounded by a cloud of gluons and quark-antiquark pairs. The effective 
mass of a constituent quark is dynamically generated and is due to the cloud of 
gluons and pairs, surrounding the valence quark. In a good approximation 
the nucleon mass is the sum of the masses of the three constituent quarks.

The internal structure of the nucleons, measured in the deep inelastic 
scattering, can then be traced back to the internal structure of the constituent 
quarks, which we denote by the capital letters {\em U}, {\em D} and {\em S}. The 
internal structure of the {\em U}-quark is given by quark and gluon distribution 
functions:
\begin{eqnarray}
u(x) ~,~ \bar{u}(x) ~,~ d(x) ~,~ \bar{d}(x) ~,~ s(x) ~,~ \bar{s}(x) ~,~ G(x)
\end{eqnarray}
Thus a constituent {\em U}-quark depends on 7 functions (the heavy quarks c, b, 
t are neglected). The constituent {\em D}-quark is obtained after the 
interchange of {\em u} and {\em d}. 

The proton consists of two {\em U}-quarks and one {\em D}-quark: $P= (UUD)$. The 
current quark distribution functions of the proton are given by:
\begin{eqnarray}
u_{p}(x)=2~ u(x) + d(x) \nonumber \\
d_{p}(x)=2 ~d(x) + u(x) \nonumber \\
\bar{u}_{p}(x)=2~ \bar{u}(x) +\bar{d}(x) \\
\bar{d}_{p}(x)= 2~ \bar{d}(x) + \bar{u}(x) \nonumber \\
s_{p}(x)=3~ s(x)\nonumber ~~~~~~~~~ \\
\bar{s}_{p}(x)= 3~ \bar{s}(x)\nonumber ~~~~~~~~~ \\
g_{p}(x)=3~ G(x) \nonumber ~~~~~~~~
\end{eqnarray}
The current quark distributions of the {\em U}-quark have to obey the following 
sum rules:
\begin{eqnarray}
\int\limits^{1}_{0}  dx(u -\bar{u}) = 1  \nonumber \\
\int\limits^{1}_{0}  dx(d -\bar{d}) = 0 \\
\int\limits^{1}_{0}  dx(s -\bar{s}) = 0 \nonumber
\end{eqnarray}
Using the relations (4), we find for the current quark distributions of the 
{\em U}-quark:
\begin{eqnarray}
u(x) = \frac{1}{3}~ (2 u_{p}(x) - d_{p}(x)) \nonumber \\
\bar{u}(x) = \frac{1}{3}~ (2 \bar{u}_{p}(x) - \bar{d}_{p}(x)) \nonumber \\
d(x) = \frac{1}{3}~ (2 d_{p}(x) - u_{p}(x)) \nonumber \\
\bar{d}(x) = \frac{1}{3}~  (2 \bar{d}_{p}(x) - \bar{u}_{p}(x)) \\
s(x) = \frac{1}{3}~  s_{p}(x) \nonumber ~~~~~~~~~~~~~~~\\
\bar{s}(x) =\frac{1}{3}~  \bar{s}_{p}(x) ~~~~~~~~~~~~~~~\nonumber \\
G(x) =\frac{1}{3}~ g_{p}(x) ~~~~~~~~~~~~~~~\nonumber
\end{eqnarray}
The sum of the contributions of the quarks and antiquarks to the nucleon 
momentum is about 45\% (ref.6):
\begin{equation}
\int\limits^{1}_{0} x[(u_{p}+\bar{u}_{p})+(d_{p}+\bar{d}_{p})+
(s_{p}+\bar{s}_{p})]dx \cong 0.45
\end{equation} 
It follows that the contributions of the gluons to the nucleon momentum must 
be about 55\%:
\begin{equation}
\int\limits^{1}_{0} xg_{p}(x)dx \cong 0.55
\end{equation} 
A constituent quark contributes 33\% to the momentum of a nucleon. Thus for the 
distribution functions of the constituent quarks {\em u}, {\em d}, etc. we find:
\begin{eqnarray}
\int\limits^{1}_{0} x[u+\bar{u}+d+\bar{d}+s+\bar{s}]dx \cong 0.15 \nonumber \\ 
\int\limits^{1}_{0} xG(x)dx \cong 0.18~~~~~~~~~~~~~~~~~~~~~~~~
\end{eqnarray} 
If we would know the six quark and antiquark distribution functions of the 
proton in detail, we would be able to determine the six distribution functions 
of a constituent quark. However the distribution functions of the proton are 
not known in detail, in particular the small strange quark distribution function. 
We shall assume that we can neglect the strange quarks in the nucleon and 
set the strange distribution function equal to zero. We assume, as expected in 
QCD, that the {\em u} and {\em d} antiquarks are the same: $\bar{u}_p=\bar{d}_p$. 
Then we can express the distribution functions of the proton as follows: 
\begin{eqnarray}
u_p=2u + d \nonumber \\
d_p=2d + u \\
\bar{u}_p=\bar{d}_p ~~~~~\nonumber 
\end{eqnarray} 
Thus the quark distribution functions of the {\em U}-quark are given by:
\begin{eqnarray}
u=\frac{1}{3} (2u_p - d_p) \nonumber \\
d=\frac{1}{3} (2d_p - u_p)  \\
\bar{u}=\bar{d}~~~~~~~~~~~~~~\nonumber 
\end{eqnarray} 
The distribution function {\em d} of a {\em U}-quark is expected to be equal to 
the antiquark distribution $ \bar {d}= \bar{u}$. They can be neglected for 
$x > {0.2}$. Thus we expect that the function {\em d} vanishes for $x > 0.2$. 
Within the experimental errors the function {\em d} vanishes.

We conclude that the constituent quarks appear as bound states of current 
quarks, antiquarks and gluons. They act as the constituents of the hadrons, 
but have their own substructure. The substructure of the nucleons, observed in 
the deep inelastic scattering, is in reality the substructure of the constituent 
quarks.

Subsequently we consider the spin structure of the constituent quarks and relate 
it to the spin structure of the nucleon. The spin vector $S_{\mu}$ of a proton 
is given by the matrix element of the axial-vector current: 
\begin{eqnarray}
2 M S_{\mu} = \langle p,s \vert \bar{\psi} \gamma_{\mu} \gamma_5 \psi \vert
p,s \rangle
\end{eqnarray}
Information about the quark ``spin content'' of the proton is obtained by the 
quark axial charges $\Delta q$:
\begin{eqnarray}
2 M S_{\mu} \Delta q = \langle p,s \vert \bar{q} \gamma_{\mu} \gamma_{5} q \vert 
p,s \rangle \\
(q: quark field) ~~~~~~~~\nonumber
\end{eqnarray}

The axial charges $ \Delta u, \Delta d $ and $ \Delta s $ can be written as 
linear combinations of the isovector, SU(3) octet and SU(3) singlet charges:
\begin{eqnarray}
g_{A}^{(3)}=\Delta u -\Delta d  ~~~~~~~~~  \nonumber \\
g_{A}^{(8)}=\Delta u +\Delta d -2 \Delta s   \\
g_{A}^{(0)}=\Delta u +\Delta d  + \Delta s ~ \nonumber
\end{eqnarray}
In the quark model $\Delta q $ is interpreted as the amount of spin, carried by 
the quarks and antiquarks of the flavor $ q $.
\medskip \\
The sum 
\begin{eqnarray}
g_{A}^{(0)}=\triangle \Sigma=\triangle u +\triangle d + \triangle s 
\end{eqnarray}
has been measured (ref.7). It is much smaller than expected in the quark model. 
\begin{eqnarray}
\triangle \Sigma\cong 0.30\pm 0.1
\end{eqnarray}

The nucleon spin can be decomposed into a quark contribution $\triangle \Sigma$, 
a gluon contribution $\triangle G$ and an orbital contribution $\triangle L$:
\begin{eqnarray}
\frac{1}{2}=\frac{1}{2} \triangle \Sigma+ \triangle G +\triangle L
\end{eqnarray}
If we take $\triangle L = 0 $, $ \triangle  G $  would have to be about 0.35.

We introduce the spin-dependent distribution functions $u_{+}, u_{-}$, etc. of 
the constituent quarks. The index ``+'' or ``-'' denotes the helicity of the 
corresponding quark or antiquark in a polarized {\em U}-quark with positive 
helicity. Especially we consider the integrals:
\begin{eqnarray}
\int\limits^{1}_{0} dx[(q_{+}+\bar{q}_{+})-(q_{-}+\bar{q}_{-})]=I_q \\ \nonumber
(q=u~,~d~,~s)~~~~~~~~~~~~~~~~~~~~~~~~
\end{eqnarray}
The difference $(I_u  - I_d)$ is the analogue of the Bjorken sum rule for 
the constituent quarks: 
\begin{eqnarray}
I_u-I_d=g_a
\end{eqnarray}
$g_a$: axial vector coupling constant, given by the isotriplet axialvector 
current. 

A polarized constituent {\em U}-quark depends in particular on the four 
functions $u_{+}, u_{-}, G_{+}$ and $G_{-}$. If we would identify the 
constituent and current quarks and use a SU(6) wave function, we would have 
the unrealistic result: 
\begin{eqnarray}
u_{+}= \delta(1-x)~~~~~~~~ \nonumber \\ 
u_{-}=G_{+}=G_{-}=0
\end{eqnarray}
The {\em U}-matrix element of the axialvector current $\langle U \vert \bar{u} 
\gamma_{\mu} \gamma_{5}u \vert U\rangle$ determines the integral $I_u$. 
In quantum chromodynamics the axial vector current is not conserved, if the
quark masses are zero, due to the gluon anomaly (ref.2):
\begin{eqnarray}
\partial^{\mu}(\bar{q} \gamma_{\mu} \gamma_{5} q )=\frac{g^{2}}{16 \pi ^{2}}
\tilde{G}G~~~~~~ \nonumber \\ 
\tilde{G}G=\sum \limits_{a} \tilde{G}^{\mu \nu }_{a}G^{a}_{\mu \nu }~~ \\
\tilde{G}^{\mu \nu }_{a}=\frac{1}{2} \varepsilon_{\mu \nu \alpha \beta } 
G_{a}^{ \alpha \beta }~~~\nonumber
\end{eqnarray}
The isotriplet matrix element $
\langle U \vert (\bar{u} \gamma_{\mu} \gamma_{5}u - \bar{d} \gamma _{\mu}
\gamma_{5}d) \vert U\rangle $ determines the integral $(I_u-I_d)$.
The corresponding isosinglet matrix element is given by the integral:
\begin{eqnarray}
\int\limits^{1}_{0} dx[(u_{+}+\bar{u}_{+}-u_{-}-\bar{u}_{-})+(d_{+}+
\bar{d}_{+}-d_{-}-\bar{d}_{-})]=\sigma
\end{eqnarray}
The number $\sigma$ can be interpreted as the contribution of the 
{\em u} and {\em d} quarks to the {\em U}-spin.

For the sum $\sigma$ we expect the same suppression as for $\Delta\Sigma$:
\begin{eqnarray}
\sigma\cong 0.30\pm0.01 \nonumber
\end{eqnarray}
Note that in the SU(3) limit we would have 
\begin{eqnarray}
d_{+}=s_{+}~,~d_{-}=s_{-}~,~\bar d_{+}=\bar s_{+}~,~ \bar d_{-}=\bar s_{-} 
\nonumber
\end{eqnarray}

In a naive model, in which there is no difference between constituent and 
current quarks, a {\em U}-quark with positive helicity would consist only of a 
{\em u }-quark with positive helicity. Thus we would have:
\begin{eqnarray}
\int\limits^{1}_{0} dx~u_{+}=1
\end{eqnarray}
\begin{eqnarray}
d_{+}=d_{-}=\bar d_{+}=\bar d_{-}=u_{-}=\bar{u}_{+}=\bar u_{-}=0, ~~~~~~
~~~~~~~~~~~~~~~~~~~~~~~~~~~~~~~~~~~~~~~~\nonumber
\end{eqnarray}
and we obtain:
\begin{eqnarray}
g_a=\sigma=+1
\end{eqnarray}
Using a SU(6) wave function for the nucleon, we then find 
\begin{eqnarray}
\frac{G_A}{G_V}=5/3 
\end{eqnarray}
($G_A$: axialvector coupling constant, $G_V$ :vector coupling constant)
\medskip \\
In the experiments one finds 
\begin{eqnarray}
\frac{G_A}{G_V}=1.2695\pm 0.0029
\end{eqnarray}

The SU(6) prediction is based on the assumption that the three quarks do not 
have orbital angular momentum({\em s}-wave). We attribute the difference between 
the SU(6)-prediction (25) and the experimental result (26) to contributions from 
angular momenta ({\em p}-wave etc.).

The reduction (ratio of the observed value of $\frac{G_A}{G_V}$ and the 
theoretical value) is about 24\%. The same reduction is expected for 
$\bigtriangleup\Sigma$, i.e. we would expect $\bigtriangleup\Sigma \approx0.76$.
In reality $\bigtriangleup\Sigma$ is about 0.30, i.e. there is a further 
reduction of 60\% from 0.76 to 0.30. In our constituent quark model the some 
reduction should present in the constituent quarks: $\sigma\approx0.40$.

This reduction should be the result of the gluon anomaly. 
The isotriplet matrix element $g_a$ is not affected by the gluon anomaly.

The integral
\begin{eqnarray}
\int\limits^{1}_{0} dx[(d_{+}+\bar{d}_{+}-d_{-}-\bar{d}_{-})\cong\frac{1}{3}
(\sigma-g_a)
\end{eqnarray}
is non-zero due to the gluon anomaly. It is given by the matrix element 
$\langle U\vert \bar {d}\gamma_\mu \gamma_5 d \vert U \rangle$. In the naive 
quark model this matrix element vanishes. We find, taking
$\sigma\cong 0.40\pm0.13$:
\begin{eqnarray}
\int\limits^{1}_{0} dx[(d_{+}+\bar{d}_{+}-d_{-}-\bar{d}_{-})\cong\frac{1}{3}
(\sigma-g_a)\cong - 0.2\pm0.04 \nonumber \\ 
\int\limits^{1}_{0} dx[(u_{+}+\bar{u}_{+}-u_{-}-\bar{u}_{-})=\frac{1}{3}(\sigma
+2 g_a)\cong 0.80\pm0.04
\end{eqnarray}
Due to the QCD anomaly $\bar {q} q$-pairs are created inside the {\em U}-state. 
The pairs are polarized and reduce the contribution of the quarks to the 
{\em U}-spin. According to the {\em d}-integral given above the sum 
$(d_{-}+\bar{d}_{-})$ must be different from zero, but the sum 
$(d_{+}+\bar{d}_{+})$ might vanish. The QCD anomaly does not distinguish between 
{\em u} and {\em d} quarks . Thus $\bar{u}u$ pairs and $\bar{d}d$ pairs are 
created with equal strength, and we expect $d_{-}=\bar d_{-}=\bar u_{-}=u_{-}$. 

A constituent {\em U}-quark can be viewed as a valence {\em u}-quark, surrounded 
by $\bar{q}q$-pairs, which reduce the spin contribution of the quarks 
significantly. It has spin $\frac{1}{2}$:
\begin{eqnarray}
\frac{1}{2}=\frac{1}{2}\cdot \sigma +\Delta G
\end{eqnarray}
($\Delta G$: gluon contribution to the spin of the constituent quark)\medskip \\
Using the quoted value of $\sigma$, we can calculate $\Delta G$:
\begin{eqnarray}
\Delta G=\frac{1}{2}(1-\sigma)\cong 0.30\pm 0.06
\end{eqnarray}

In this calculation  we have used the SU(3) symmetry. Any SU(3)-breaking 
would reduce the integral over the strange quark densities. If we neglect the 
strange quarks$(s_{+}=s_{-}=\bar s_{+}=\bar s_{-}=0)$, we would find:
\begin{eqnarray}
\Delta G\cong0.35\pm0.06,
\end{eqnarray}
thus the gluonic contribution would increase slightly.

Using eq.(30), we find that about 40\% of the spin of a constituent quark is 
provided by the quarks, and 60\% is provided by gluons. A polarized constituent 
quark consists of a polarized valence quark, surrounded by polarized 
quark-antiquark pairs and by a cloud of polarized gluons, which provide a large 
part of the spin.

In the simple SU(6) model about 24\% of the nucleon spin is due to
orbital angular momenta of the three quarks, and 76\% is due to the spins
of the quarks. The nucleon spin 1/2 is given by:
\begin{eqnarray}
\frac{1}{2}=(\frac{1}{2}\times0.76)+(\frac{1}{2}\times0.24)
\end{eqnarray}
In our model we find:
\begin{eqnarray}
\frac{1}{2}=\frac{1}{2}\Delta\Sigma+\Delta G+\Delta L \nonumber~~~~~~~~~~~~~~~~~
~~~~~~~\\
=(\frac{1}{2}\times0.30)+(\frac{1}{2}\times0.46)+(\frac{1}{2}\times 0.24)
\end{eqnarray}
About 24\% of the nucleon spin is due to orbital angular momenta, about 46\%
is due to the spins of the gluons inside the constituent quarks, and only 30\% 
is due to the spins of the quarks.

\end{document}